\documentclass[aps,prl,twocolumn,showpacs,epsfig,epsf,amssymb,floatfix,reprint]{revtex4-1}
\usepackage{graphicx}
\usepackage{amssymb}
\usepackage{amsmath}
\usepackage{subfigure}
\usepackage{booktabs}
\usepackage{float}
\usepackage{longtable}
\begin{document}
\newcommand{\beginsupplement}{%
        \setcounter{table}{0}
        \renewcommand{\thetable}{S\arabic{table}}%
        \setcounter{figure}{0}
        \renewcommand{\thefigure}{S\arabic{figure}}%
     }


\title{ Origin of spatial organization of DNA-polymer in bacterial chromosomes.
}

\author{Tejal Agarwal$^{1}$, G.P. Manjunath$^2$, Farhat Habib$^3$, Apratim Chatterji$^{1,4}$}
\email{apratim@iiserpune.ac.in}
\affiliation{
$^1$ IISER-Pune, 900 NCL Innovation Park, Dr. Homi Bhaba Road,  Pune-411008, India.\\
$^2$ IISER Mohali, Knowledge city, Sector 81, SAS Nagar, Manauli-140306, India.\\
$^3$ Inmobi, Cessna Business Park, Outer Ring Road, Bangalore-560103, India.\\
$^4$ Center for Energy Science, IISER-Pune,  Dr. Homi Bhaba Road,  Pune-411008, India.
}

\date{\today}
\begin{abstract}
In-vivo DNA organization at large length scales ($\sim 100nm$) is highly debated and polymer modelshave proved useful to understand the principle of DNA-organization. 
Here, we show that $<2$\% cross-links at specific points in a ring polymer can lead to a distinct spatial
organization of the polymer. The specific pairs of cross-linked monomers were extracted from contact
maps of bacterial DNA.
We are able to predict the structure of 2 DNAs using Monte Carlo simulations
of the bead-spring polymer with cross-links at these special positions. Simulations with cross-links at random
positions along the chain show that the organization of the polymer is different in nature from the previous
case.
\end{abstract}

\maketitle

It is established that DNA-polymer is not a random coil in either bacterial cells \cite{laub,joyeux,cagliero} 
or in eukaryotic cells \cite{aiden,Wendy,sexton,tjong}.
Experimental methods such as CCC (chromosomal conformation capture) which was then further developed as 5C and then Hi-C
 have consistently shown the presence of  topologically associated domains (TADs) in the contact maps (C-maps)
of DNA-chains \cite{dixon,kremer,dekker}. The Hi-C technique gives us the C-map which is the map of  frequencies that a segment 
of the DNA chain (say $i$) is found in spatial proximity to another segment (say $j$) for all combinations $i,j$ of
segments along the contour length of the DNA-polymer.  TADs are patches  in C-maps 
which indicate that some segments of the chain (at 1 mega-base pair(BP) to 1 kilo-BP resolution),
are found spatially close to other particular segments with higher frequencies compared to the rest of the segments.

The ds-DNA is stiff at length scales of $1$nm but can be considered to be a flexible
chain at length scales beyond $100$nm \cite{rob} . The persistence length $\ell_p$
of a naked DNA is 150 Base Pairs (BP) $\equiv 50$ nm \cite{colby} and the value of  $\ell_p$ in vivo is debated \cite{Maeshima}. 
Since, the resolution of Hi-C experiments are well above this length scale \cite{aiden,laub}, there has a focussed attempts
in the last few years trying to understand the DNA organization and in particular origin of formation of TADs from the principles of polymer
physics \cite{nicodemi,marenduzzo,mirny_cell_reports,mirny3,rosa}. A series of studies indicate that TADs in eukaryotic cells are indicative of fractal globule organization of the polymer
(as opposed to equilibrium globule) \cite{aiden,mirny2}. Recently, more detailed polymer models with either different lengths of 
loops or with many distinct (coarse-grained) diffusing binder molecules which cross-link different segments of the chain  
have reproduced TADs of sections of a particular eukaryotic DNA by performing optimizations in
multi-parameter space. Distinct kinds of binder molecules link correspondingly distinct monomers (DNA-segments) 
along the chain, and the optimization parameters include the number of distinct kind of binders/monomers as well
as the position and number of  distinct monomers as well as diffusing cross-links along the contour \cite{mariano,geoffrey,mirny_cell_reports,anton}.

We propose a much simpler model for shorter bacterial DNAs and ask a more general question: 
Does fixed cross-links (CLs) at a few specific positions along the polymer chain contour organize the  
polymer into a particular architecture? 
If so, can we predict the global shape/structure of the DNA polymer and does it reproduce the C-map or at least
parts of it?  The position of the cross-links are chosen by using the C-map to  identify
the highest frequencies of two segments to be in spatial proximity. We cross-link a minimal number of these segments,
and then computationally cross-check if the other segments of the polymer get localized in space and with respect 
to each other.  Of course the chain can fluctuate due to thermal fluctuations but maintaining the architecture. 
We then compare this polymer organization with the organization obtained when a ring-polymer (most bacterial DNAs are 
ring polymers) has an equal number of CLs at randomly chosen positions along the chain contour.
We choose $10$ different realizations of randomly positioned CLs. On comparing we see that nature chooses the position of CLs
carefully such that the architecture of the DNA-polymer is well organized in a manner very distinct from what 
is obtained for a polymer with random CLs. 
   
We investigate the organization of two bacterial DNAs, {\em E. Coli} and {\em C. Crescentus}.
Each has a single chromosome of length $4$ Mega-BP: we choose to 
work with shorter bacterial DNA with just one chromosome and no nucleus wall. 
DNA is modeled as a flexible bead spring ring polymer (both bacterial chromosomes are ring polymers) 
with a harmonic spring potential $\kappa(r-a)^2$ acting between  neighbouring beads; the choice of 
$a=1$ ($\sim 100$ nm) sets the length scale of the problem. 
The excluded volume(EV) of the beads are modelled by suitably truncated purely repulsive 
Lennard-Jones potential with $\sigma=0.2a$. 
The E.coli and {\em C.Crescentus} DNAs have $4642$ and $4017$ kilo-BPs,
which we model by $4642$ and $4017$ monomers, respectively. The naked DNA Kuhn segment has just 300 BPs
\cite{colby} whereas a bead represents 1000 BPs. The effects of DNA coiling around histone-like proteins occurs at smaller
length scales; longer range effects due  supercoiling, presence of plectonemes etc. should show
up in the C-map and their effects gets incorporated as cross-links at the length scales we consider.
Moreover, bacterial DNA
occupy $15-25$\% of cellular volume, so we choose to ignore confinement effects, if any. Instead, we fully focus 
on the role of CLs in the organization of the polymer. The introduction of CLs between segments of the chain in our model
can be justified due the presence of DNA-binding proteins  
which are present in bacterial cells (as well as higher eukaryotic cells) \cite{joyeux,dixon,phillips,ohlsson}.

We model the cross-links between two segments of the DNA-polymer by a harmonic potential [$\kappa(r-a)^2$, where $\kappa=200k_BT/a^2$] between 
two monomer beads. The two ``cross-linked'' monomers (CLs) are typically well separated along 
contour of the model polymer. 
We cross-link pair of monomers if they are found spatially close above a certain frequency in C-maps.
By lowering the frequency cutoff, we can have more cross-linked monomers. For details, refer \cite{condmat}. 
So, we take $47$ or $159$ CLs for {\em E. Coli}. For {\em C.Crescentus} we take $49$ or $153$ CLs which we refer
as BC-1 and BC-2, respectively. The list of cross-linked monomers are listed in Table S\ref{tab:table1}.
From the table one observes that a pair of neighbouring monomers along the chain contour can get cross-linked to 
another pair of neighbouring monomers, hence the number of {\em actual} 
CLs  are fewer (refer Table \ref{tab:table1} for examples and detailed explanations) . 
Removing such over-counts, there are $26$ and $60$ {\em effective} CLs for {\em C.Crescentus}. 
For {\em E. Coli} we have $27$ and $82$ effective CLs, CL-list and other details are in \cite{condmat}.     
We also investigate large scale organization of the chain when we have a set of CLs, where pairs of monomers are
chosen randomly and cross-linked.
A set of $26$ and $60$ CLs at random positions in a ring of $4017$ monomers is referred as RC-1 and RC-2, respectively.

\begin{figure}[!tbh]
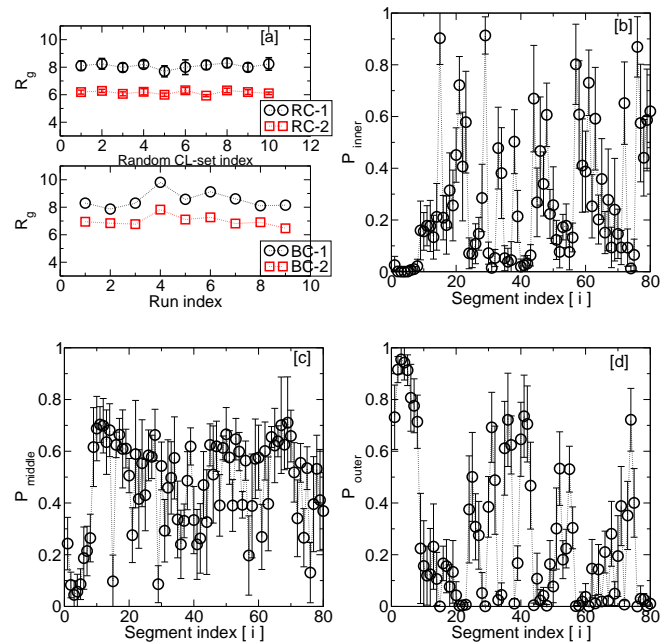

\includegraphics[width=0.49\columnwidth]{R_g.eps} 
\includegraphics[width=0.49\columnwidth]{caul_cm_contact4_inner.eps}
\vskip0.4cm 
\includegraphics[width=0.49\columnwidth]{caul_cm_contact4_middle.eps}
\includegraphics[width=0.49\columnwidth]{caul_cm_contact4_outer.eps} 
\caption{\label{fig1} (a) shows that the $R_g$ of {\em C.Crescentus} DNA for  
9 runs, each starting from  independent initial conditions for BC-1 and BC-2 sets of CLs. It 
also shows $R_g$ data for random set of CLs: RC-1 and RC-2 for 10 different choices of random cross-links sets, 
set-index is on x-axis.  The error bar shown for  $R_g$ is the standard deviation (SD) calculated from $9$ indepedent 
runs for a particular set of CLs. 
Subplots (b),(c) and (d) show the probability $P_{inner},\,P_{middle}, \,P_{outer}$ of finding a segment $i$ in the inner, middle
and outer regions, respectively, of the DNA-coil for BC-2. The error bar is the SD calculated from $9$ independent runs.
} 
\end{figure}

\section{Results} 
To establish that  sets of bio-CLs lead to a particular organization of the polymer,
we start from $9$ different initial configurations of a ring polymer, and allow the chain to {\em equilibrate} 
using Monte Carlo (MC) simulations using Metropolis algorithm.
After equilibration (inherently non-equilibrium biological systems at a certain stage of their cell cycle 
can be thought of to be in a state of local equilibrium), we compare statistical quantities  which provide us evidences 
about structure and conformations of polymer chains.  If we get similar structure from all $9$ runs, we could claim 
that CLs lead to particular sets of conformation at large length-scales. 
We take care to choose the $9$ different initial configurations of polymers in ways that 
we ensure that the cross-linked monomers are at very different relative positions with respect to each other, 
moreover, distance between them can be much larger than their equilibrium distance $a$. 
Thus the initial potential energy of the system will be high (Fig.S3), and
as the system relaxes due to the presence of CLs through MC moves, the average energy of the polymers in each run 
should have nearly the same value at the end of equilibration run. 
After equilibration, the polymer explores phase space over $12$ million iterations in each run to calculate 
statistical quantities with thermal energy scale $k_BT=1$.  
Small EV of the beads allows chains to cross each other, moreover we take a large MC displacement 
of $1.2\sigma$ in every 100 steps. These help in releasing any artificial 
topological constraints induced by intial configuration. Chain crossing is justified due to the activity of topoisomerase II.

We repeat these calculations using RC-1 and RC-2 CLs and compare polymer organization with those obtained using
BC-1 \& BC-2. To firmly establish that the BC-1 and BC-2 set of CLs lead
to an organization of the ring polymer which is very distinct from the organization achieved with random CLs, 
we choose $10$ independent sets of random CLs, then for each set of CLs gave $9$ runs starting from $9$ independent initial
positions. After equilibration, we compare the differences in the large-scale organization. For each random CL-set,
RC-1 CLs are a subset of RC-2 CL set.  We show later that the reason 
for the distinct organizaton of the chain with BC-2 is in turn the very distinct spatial organization of the CLs themselves
in space, which in turn comes from choice of monomers which are cross-linked.    

The primary problem is how to identify large scale organization of a single floppy polymer chain and come
up with a prediction of relative position of different segments, when rapid comformation changes
are inherent in the system. Quantities like pair correlation function $g(r)$ between monomers are insufficient as we 
would like more individualized information about arrangement of different segments of the DNA. 
We use the following four quantities to identify the global organization of {\em C.Crecentus} DNA-polymers;
similar detailed data for {\em E.coli} is given in \cite{condmat}

1. We estimate the radius of gyration $R_g$ of the DNA-polymer of {\em C.Crescentus}. 
The value of $R_g$ obtained is $\approx 8a$ from all the $9$ runs for BC-1, and 
$\approx 6.5a$ for BC-2. Data is shown in Fig \ref{fig1}a. In contrast, $R_g$ decreases from $8a$ to
$6a$, when number of CLs are increased from RC-1 to RC-2 for each set of random CLs. 
The decrease is more significant for random CLs,
as would be expected for a polymer chain with more constraints.   
A smaller relative decrease in $R_g$ as we change from BC-1 to BC-2 compared 
to the change from RC-1 $\rightarrow$ RC-2 is the first indication of distinct organization of the coil with bio-CLs.
A ring polymer with $4017$ monomers without any CLs has $R_g$ value of $11$.

2. We divide the polymer into segments of $50$ monomers each, and identify whether the center of mass (CM) of 
each segment is to be found in the inner, middle or outer section of the coil. Thereby the polymer 
has $80$ segments. 
We define a segment to be in inner/middle/outer section if the distance $r$
of segment's CM  from the CM  of the  coil is ($r<5a$),\,/($5a<r<9a$),\,/($r>9a$), respectively. 
If a segment $i$ is found to be in the same section in all the $9$ independent runs,   
we can claim that all $9$ chains are similarly organized.  
Data in Fig.\ref{fig1} b,c,d, confirms and validates the above claim.  
As we see in Fig.\ref{fig1}b, the same segments are found inside of the coil with higher probability,
some are more likely to be found in the middle region, and the rest in the outer region.
However, the values of probabilities $P_{inner},P_{middle},P_{outer}$ for a segment  statistically fluctuate across runs.

\begin{figure}[!hbt]
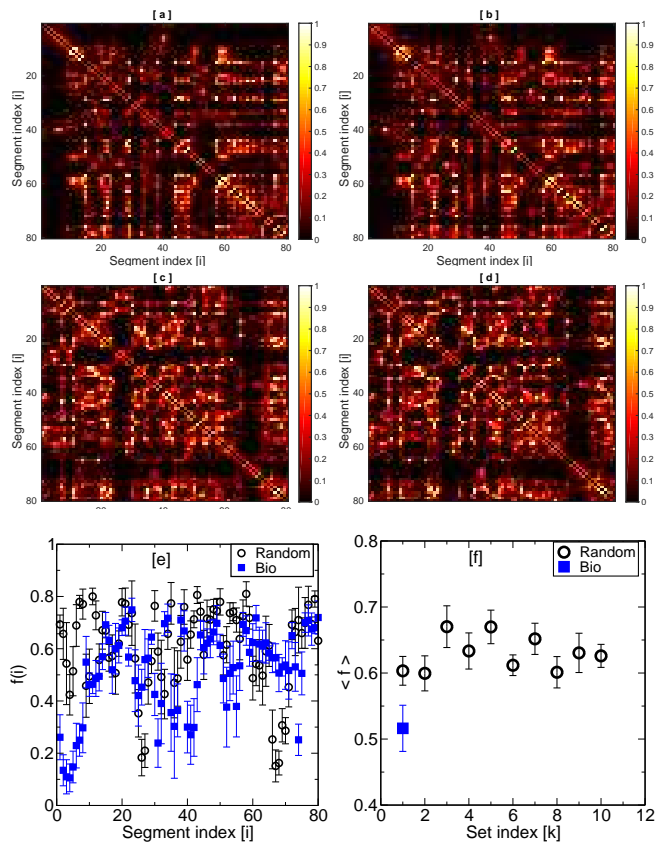

\includegraphics[width=0.49\columnwidth]{bio_cc_i1_cp4_cm_dom_corr1_l1.eps} 
\includegraphics[width=0.49\columnwidth]{bio_cc_i3_cp4_cm_dom_corr1_l1.eps} \\

\includegraphics[width=0.49\columnwidth]{r_bio_cc_i1_cp4_cm_dom_corr1_l1.eps} 
\includegraphics[width=0.49\columnwidth]{r_bio_cc_i3_cp4_cm_dom_corr1_l1.eps} \\

\includegraphics[width=0.49\columnwidth]{diff_cm_dom_corr1.eps}
\includegraphics[width=0.49\columnwidth]{diff_cm_dom_corr1_avg.eps}
\caption{\label{fig2} 
The colormaps give the probability of finding the CM of segment $i$ and 
$j$ within a distance cutoff of $R^C = 5a$ for a DNA polymer with cross-links. 
Figure (a),(b) correspond to two independent runs starting with different initial configurations for BC-2.
and (c),(d) correspond to the runs with different intial conditions for 
RC-2. Data for additional runs for BC-1, BC-2 and RC-1, RC-2 are given in Fig.S1. 
In plot (e) the x-axis shows segment index and y-axis shows [$f(i)$] which is the number of pixels with probability
$p(i,j) > 0.05$ (non-dark pixels) divided by the total number of segments in subplots (a),(b),(c),(d) for BC-2 and RC-2
respectively. Error bars shows the fluctuations in the value of $f(i)$ across independent $9$ initial conditions.
Subplot (f) shows the average value of $f(i)$ ($\langle f \rangle$)for $10$ random CL-sets and $1$ bio-CL set BC-2.
} 
\end{figure}

3. Instead of calculating $g(r)$, we aim to identify which segments (say $i$) are near other 
segments (say $j$) with higher probability. 
We can calculate this probability for each pair of segments
and show this in a color-map, where  both the axes represent segment indices and the color
of each pixel denotes  the probability that the CM of segments $i$ and $j$ 
are within distance of $R^C< 5a$. 
We show data from two runs for BC-2 in Fig.\ref{fig2} (a,b). 
For comparison we also show data RC-2 in  Fig.\ref{fig2} (c,d) , respectively. 
Colormaps of bio-CLs and  random-CLs  in Fig.\ref{fig2} show that there 
is higher probability (color red) of finding only certain segments near others, and some segments are 
never found  in proximity of certain other segments (color black).  
This indicates a certain degree of organization of segments. 
Large fluctuations in the conformation of polymer would result in a colormap which would be predominantly dark, indicating that 
there is almost equal (and small) probability of different segments to be near each other(Fig.S1). 
Moreover, colormaps from independent runs and same CL set show statistically similar patterns of red and dark pixels:
this implies the same organization of segments in independent runs.

\begin{figure}[!hbt]
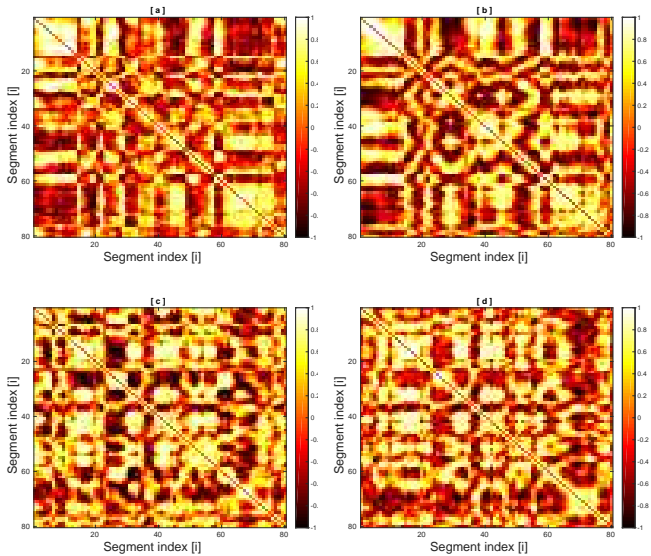

\includegraphics[width=0.49\columnwidth]{bio_cc_i1_cp4_near_contacts1_l1.eps}
\includegraphics[width=0.49\columnwidth]{bio_cc_i3_cp4_near_contacts1_l1.eps}
\vskip0.4cm 
\includegraphics[width=0.49\columnwidth]{r_bio_cc_i1_cp4_near_contacts1_l1.eps} 
\includegraphics[width=0.49\columnwidth]{r_bio_cc_i3_cp4_near_contacts1_l1.eps} 
\caption{\label{fig3} 
The set of 4 colormaps give the probability of finding the angular correlations between segment $i$ and 
segment $j$ of a ring polymer. 
Figure (a),(b) correspond to independent runs of the {\em C.Crescentus} DNA-polymers with BC-2 set of CLs. 
Subplots (c),(d) are for runs with CL-set RC-2.
Colormaps from additional  runs for BC-1, BC-2, RC-1 and RC-2 are given in Fig.S2.}
 
\end{figure}

For BC-1/RC-1, we cannot identify large scale structural organization of the polymer, nor can one 
distinguish between the colormaps of BC-1 and RC-1.  In contrast, on comparing colormaps of BC-2 and 
RC-2 (Fig.\ref{fig2}) we can infer that the character of organization of the polymer is different in the two cases.   
We find large red patches separated by dark rows/columns in the colormaps for BC-2, whereas, for RC-2 colormaps the lighter pixels 
are relatively more scattered. This is further quantified in Fig.\ref{fig2}e, where we see that a chain segment is 
approached by fewer other segments for BC-2 as compared to RC-2. For each of the random CL-sets, each
segment can be near a larger number of segments as can be deduced from the higher value 
of $\langle f \rangle$ in Fig.\ref{fig2}f. This observation, coupled with the fact 
that polymer has higher value of $R_g$ and more segments in the outer region for BC-2 when compared with data 
for  RC-2(Fig.S6), implies that
certain segments have well defined neighbouring segments (more structure) as compared to polymer with RC-2. 
The neighbouring segments could be far away along the contour length but are neighbours spatially.
Thus, the positions of bio-CLs along the contour are special (not random) for BC-2, 
as these result in distinctive meso-scale organization of the DNA.

We emphasize that the colormaps of Fig.\ref{fig2} look  similar  to the C-maps 
of DNA-polymer which we use as modelling input. But the content is very different in the sense we 
obtain large scale structural correlations of the entire polymer chain  from our colormaps. 
To reiterate, C-maps give us input about the location of CLs along the polymer contour
at the length scale of monomers, our color-maps  show how various segments 
(each of 50 monomers) are organized relative to each other in space. 

4. We next focus on relative angular position of polymer segments with respect to the CM of 
the DNA globule.  For each pair of segments $i,j$ we calculate if the vectors $\vec{r_i}$ and 
$\vec{r_j}$, joining the CM of globule to the CM of the segments $i,j$, subtend an angle of more than  
$\pi/2$ radians. If the angle $\theta$ between  $\vec{r_i}$ and $\vec{r_j}$ 
is $ >\pi/2$, then we interpret that the two segments lie in opposite hemispheres, else the two segments lie 
on the same side  of the globule w.r.t the CM of the polymer.
We compute the average of the counter $\delta^{ij}_\theta$ for each pair of segments $i,j$ as follows:
for a microstate $\delta^{ij}_\theta$ is incremented by $1$ if $\cos(\theta)<0$, and decremented 
by $1$ if $\cos(\theta)<0$. We plot the value of $\langle \delta^{ij}_\theta \rangle$ $N^{ave}$ in Fig.\ref{fig3} for 
each pair of $i-j$ for BC-1,BC-2,RC-1,RC-2 CLs.  
The value of $\langle \delta^{ij}_\theta \rangle$ varies from $-1$ to $1$.

\begin{figure}[!hbt]
\includegraphics[width=0.49\columnwidth]{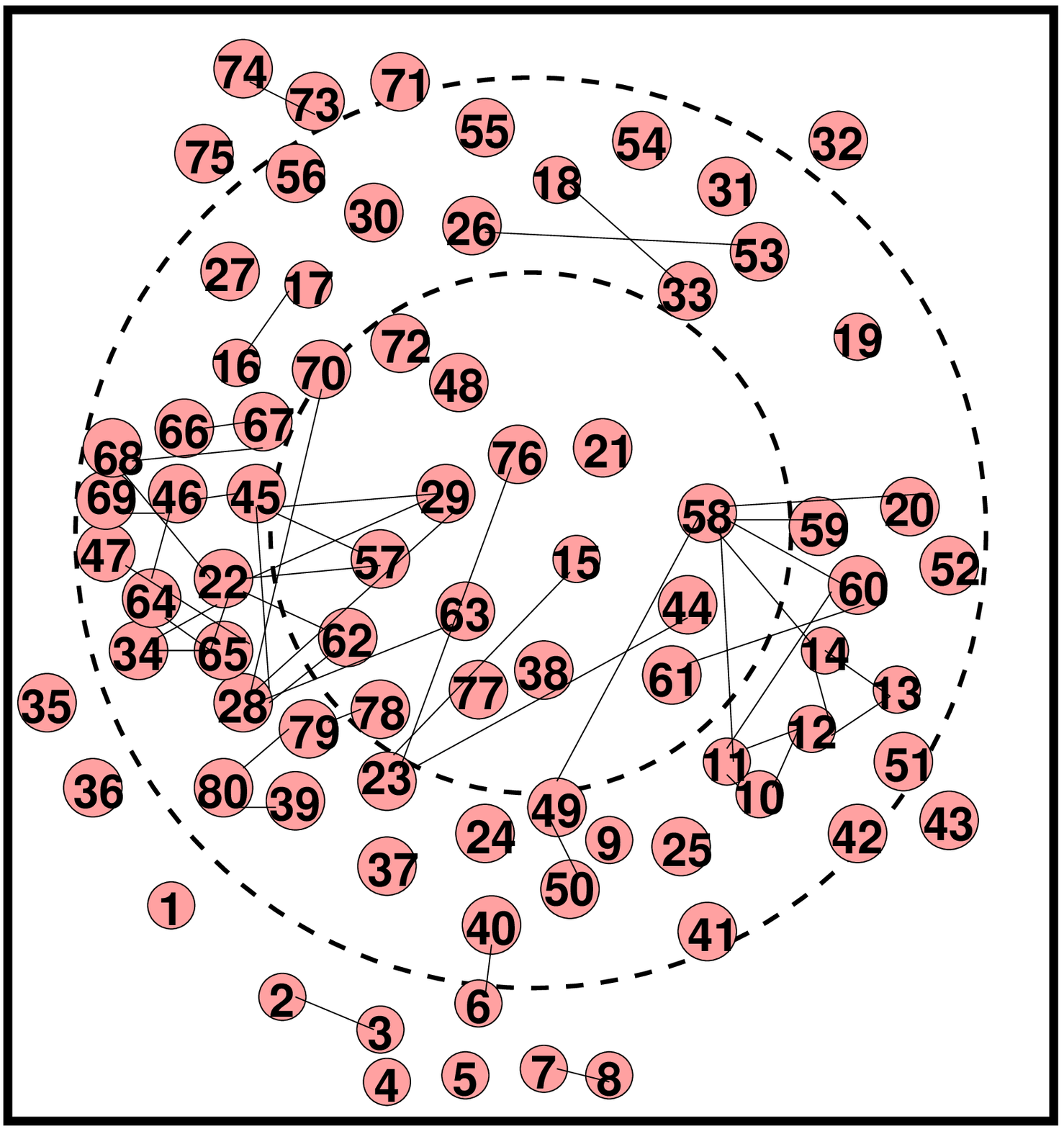} 
\includegraphics[width=0.49\columnwidth]{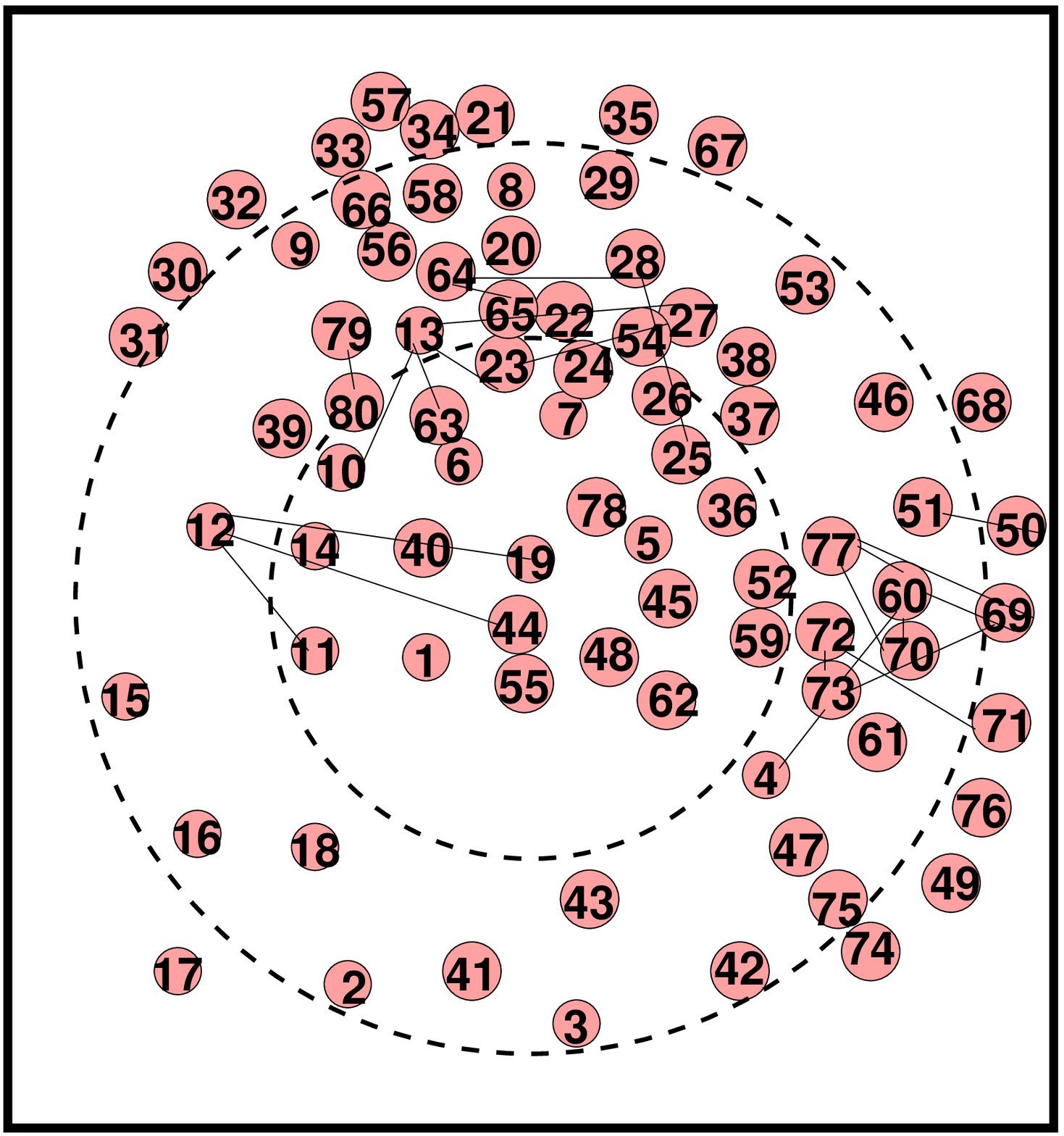} 
\caption{\label{fig4} The 2-d maps of DNA organization of {\em C. Crescentus} and {\em E.coli} bacteria above
the $100$ nm length scale using BC-2. This is obtained using the statistical data presented in this manuscript, and 
in \cite{condmat} for {\em E.Coli}. The links between different circles indicate that colormaps show these segments to
be spatially close to each other though they look separated on the map.} 
\end{figure}

The interpretation of data presented in Fig.\ref{fig3} is similar to that of Fig.\ref{fig2}.
If the pixel corresponding to segments $i,j$ is bright, then they are angularly
close w.r.t. coil CM, and a dark pixel indicates they lie predominantly  on opposite hemispheres. 
An orange pixel corresponds to the value of $\langle \delta^{ij}_\theta \rangle \approx 0$. 
If $\langle \delta^{ij}_\theta \rangle \approx 0$ then we cannot interpret their relative angular locations. 
An orange pixel does not imply  $\theta \approx  \pi/2$  because one can also get $\langle \delta^{ij}_\theta \rangle \approx 0$
if the segments  lie close to center of the coil and can rapidly change their relative positions by 
small spatial displacements. This would lead to $\langle \delta^{ij}_\theta \rangle \approx 0$.    
For BC-1 and RC-1 CL sets(Fig.S2), we again see that the patterns of bright and dark pixels 
are almost identical from independent runs. 
In contrast, the colormaps for BC-2 in Fig.\ref{fig3} (a,b) and RC-2 in (c,d) are immediately 
distinguishable. The BC-2 data  show large patches of bright pixels indicating
that adjacent segments along the contour are on the same side of the globule. The dark and bright pixels for RC-2 
is relatively more distributed/scattered. 
Furthermore, more detailed discussion on the reasons for differences  
color-maps in Fig.\ref{fig2} and \ref{fig3} for BC-2 are given in Supplementary section.

Finally, using the aggregate of all the  structural quantities presented in Figs.\ref{fig1},\ref{fig2}, \ref{fig3} we are able to piece together 
the large scale organization of DNA-polymers in a 2D map for both {\em C.Crescentus} and {\em E. Coli}(Fig.\ref{fig4}). For details see supplementary section: Construction of 2-D map.
Corresponding structural quantities/colormaps for {\em E. Coli} can be found in \cite{condmat}; the interpretation and 
conclusions are similar to what is discussed above for {\em C.Crescentus}.

\begin{figure}[!hbt]
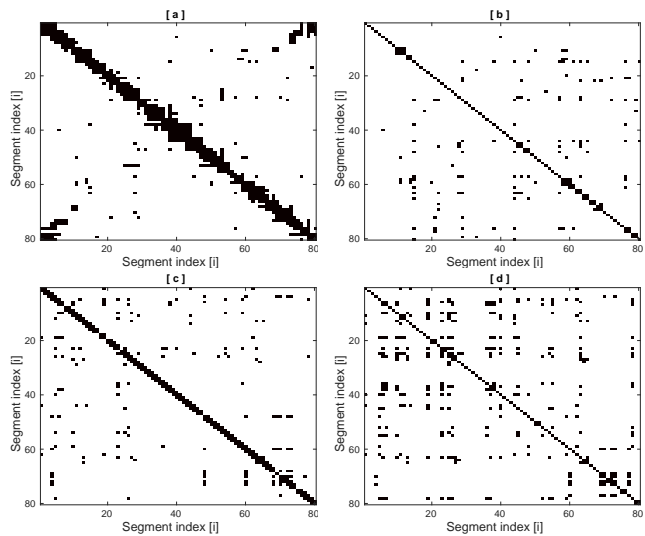

\includegraphics[width=0.48\columnwidth]{caul_contact_binary.eps}
\hskip0.05cm 
\includegraphics[width=0.48\columnwidth]{caul_sim_binary.eps} \\
\includegraphics[width=0.48\columnwidth]{ecoli_contact_binary.eps}
\hskip0.05cm 
\includegraphics[width=0.48\columnwidth]{ecoli_sim_binary.eps}
\caption{\label{fig12} 
Plots (a), (b) show binary version of 
the coarse-grained experimental C-map and color map obtained from simulations respectively, for 
{\em C. Crecentus}. Plots (c),(d) are corresponding data for {\em E.Coli}. To be able to compare easily by the eye,
we have color coded all the frequencies $f$ [probabilities $p$] with value $f> 0.0001$ [$p >0.5$] as black for experimental [simulation]
colormaps.  The corresponding actual (non-binary) coarse-grained experimental data is given in Fig.S4}
 
\end{figure}

In lieu of the future experiments which can confirm our predictions, we can use the information content of the contact map
to validate our prediction of DNA-organization. To be able to compare the positional colormaps 
obtained from our simulations to the experimental C-map, we condense the data of C-maps into a $80 \times 80$ 
matrix by suitable coarse-graining (averaging over neighbouring bins) of the frequency data of C-maps. 
We can now compare  the coarse-grained experimental C-map (which now give proximity frequencies of large sections of the DNA) 
with the highest probabilities ($>0.5$) of the color-maps (generated by our simulations) as shown in Fig. \ref{fig12}, 
these show a good match with each other.
This is by no means an obvious match: we take very few (but significant) points from the  C-map as cross-linked monomers 
in our simulations, and are able to predict from our simulations the positions of highest contact frequencies
in the coarse-grained C-map. In addition, we generate from our simulations the radial position of segements, 
angular positions of segments which is not available in the C-maps.  The difference between the coarse-grained experimental
and simulation color-map for {\em C.Crecentus} is the absence of clear high-probability diagonals in our colormap.
The plausible reason for the presence of the diagonal and off-diagonal bright patches in the experimental C-map 
could be the presence of plectonemes as proposed in \cite{laub}. The effect 
of plectonemes are not accounted for in our model as we have taken only 153 CLs which translate to only 60 effective CLs.
But using these CLs our simulation C-maps matches accurately  with the high probablity pixels for segments far separated along the contour
in the experimental C-map. Furthermore, the cutoff distance $R^C=5a$ we choose while generating 
colormaps is comparable to the $R_g \approx4a$ of 50 segments. Thereby, only few segment CMs can
occupy positions within distance $R^C$.

To summarize,  we show that the underlying mechanism of meso-scale organization
in {\em E. Coli} and {\em  C.Crescentus} DNA
involves constraints (cross-links) at specific positions along the chain contour.  
Also we predict the overall 2D architecture of the bacterial genomes.
We observe that the nature of the organization is different for polymers with the CLs taken from the experimental 
C-maps and for polymers with CL-positions chosen randomly.
Our preliminary understanding for this difference is that for bio-CLs multiple CLs get clumped together spatially (Fig.S??). As a consequence  multiple segments of the chain are pulled in together
towards the center of the coil with loops remaining on the outside. This can be reconfirmed also for {\em E.Coli} in \cite{condmat}.
In contrast for RC-2 set of CLs the CLs are scattered in space. We have also shown that
a minimal number (around $\sim 3 \%$ of monomers) of CLs are required for a polymer to get organized in a particular 
structure as we do not get any organization in the case of BC-1 and RC-1. We have checked lack of clear polymer 
organization with the number of CLs in between BC-1 and BC-2.
  

\bibliographystyle{apsrev4-1}
\bibliography{prl}
\clearpage
\begin{center}
\textbf{\large Supplementary Materials}
\end{center}
\beginsupplement
\section{List of cross-linked monomers in our simulations.}
In the following table, we list the monomers which are cross-linked to model the constraints for
the DNA of bacteria {\em C. Crecentus}. Note that for random cross links (CL) set-1 and set-2 (RC-1, RC-2)
we have fewer number of CLs, as there are fewer {\em effective} CLs.  

In particular while counting the number of independent CLs, one should 
pay special attention to the points listed below. As a consequence, $49$ CLs of BC-1 should
be counted as only $26$ independent and thereby $26$ {\em effective} CLs. Hence, we use just $26$ CLs in RC-1, when we compare organization
of RC-1 and  BC-1. Correspondingly, we have just $60$ CLs in RC-2, instead of $153$ in BC-2.  \\
\begin{itemize}
\item The rows corresponding to independent cross-links of set BC-1 are marked by $^*$, one can observe that the next row
of CLs are adjacent to the monomers marked just previously by $^*$. These cannot be counted as independent CLs.
\item The rows marked by $^+$ is not a independent CL at all, monomers $2581,2582$ and $2584$ are trivially close to
each other by virtue of their position along the contour.
\item There are some CLs marked by $^{**}$, where one monomer is linked to 2 different (non-adjacent) monomers.
\end{itemize}
Similar careful identification of effective CLs was done for BC-2 as well. 
As one can calculate, there are  $26$ and $60$ effective CLs for {\em C.Crescentus}.  
For {\em E. Coli}, the corresponding calculation of effective CLs ($27$ and $82$ for BC-1 and BC-2, respectively)
in shown in \cite{condmat}.

\begin{longtable*}{|l|l|l|l|l|l|l|l|l|l}
\hline
- & \multicolumn{2}{c|}{BC-1} &\multicolumn{2}{c|}{RC-1} &\multicolumn{2}{c|}{BC-2} & \multicolumn{2}{c|}{RC-2} \\
\hline
\midrule
Serial  & Monomer & Monomer & Monomer & Monomer & Monomer & Monomer &  Monomer & Monomer  \\
no.  &  index-1 & index-2 &  index-1 & index-2&  Index-1 & Index-2 &  Index-1 & Index-2  \\
\hline
\hline
1 & 1$^*$  & 4017  & 1  & 4017  & 1  & 4017&      1   &  4017 \\
2 & 289$^*$ & 1985 & 23 & 2743 & 289 & 1985 & 23 & 2743 \\
3 & 289 & 1986 & 2348 & 3956 & 289 & 1986 & 2348 & 3956 \\
4 & 290 & 1986 & 2602 & 3884 & 290 & 1986 & 2602 & 3884 \\
5 & 290 & 1987 & 3724 & 2295 & 290 & 1987 & 3724 & 2295 \\
6 & 468$^*$ & 564 & 2675 & 2406 & 438 & 3797 & 2675 & 2406 \\
7 & 469 & 565 & 1972 & 2520 & 468 & 563 & 1972 & 2520 \\
8 & 470 & 566 & 3779 & 1729 & 468 & 563 & 3779 & 1729 \\
9 & 470 & 567 & 3022 & 3962 & 469 & 565 & 3022 & 3962 \\
10 & 471 & 567 & 1093 & 2574 & 469 & 566 & 1093 & 2574 \\
11 & 541$^*$ & 2494 & 2739 & 3649 & 470 & 566 & 2739 & 3649 \\
12 & 541$^*$ & 2907 & 958 & 3944 & 470 & 567 & 958 & 3944 \\
13 & 541$^*$ & 2957 & 2611 & 1071 & 471 & 567 & 2611 & 1071 \\
14 & 541 & 2958 & 512 & 1466 & 540 & 955 & 512 & 1466 \\
15 & 641$^*$ & 683 & 229 & 1385 & 540 & 2494 & 229 & 1385 \\
16 & 693$^*$ & 2875 & 3206 & 679 & 540 & 2906 & 3206 & 679 \\
17 & 693 & 2876 & 2471 & 3744 & 540 & 2957 & 2471 & 3744 \\
18 & 693$^*$ & 2945 & 663 & 2183 & 541 & 2493 & 663 & 2183 \\
19 & 710$^*$ & 1890 & 1906 & 3786 & 541 & 2494 & 1906 & 3786 \\
20 & 710$^*$ & 3145 & 2029 & 1691 & 541 & 2906 & 2029 & 1691 \\
21 & 733$^*$ & 1890 & 1218 & 199 & 541 & 2907 & 1218 & 199 \\
22 & 847$^*$ & 3912 & 1478 & 2675 & 541 & 2957 & 1478 & 2675 \\
23 & 1032$^*$ & 1437 & 906 & 1605 & 541 & 2958 & 906 & 1605 \\
24 & 1032$^*$ & 3851 & 1124 & 3776 & 542 & 2907 & 1124 & 3776 \\
25 & 1033$^{**}$ & 3579$^{**}$ & 1656 & 881 & 542 & 2958 & 1656 & 881 \\
26 & 1261$^*$ & 2613 & 2183 & 332 & 542 & 2907 & 2183 & 332 \\
27 & 1369$^*$ & 2250 & - & - & 640 & 683 & 3083 & 1374 \\
28 & 1370 & 2249$^{**}$ & - & - & 641 & 683 & 2086 & 742 \\
29 & 1370 & 2250 & - & - & 693 & 2875 & 2145 & 314 \\
30 & 1393$^*$ & 3119 & - & - & 693 & 2876 & 748 & 841 \\
31 & 1398$^*$ & 3455 & - & - & 693 & 2945 & 3555 & 3745 \\
32 & 1399 & 3454 & - & - & 693 & 2946 & 1687 & 2814 \\
33 & 1399 & 3454 & - & - & 694 & 2876 & 1723 & 520 \\
34 & 1437$^*$ & 3580 & - & - & 694 & 2944 & 2480 & 2903 \\
35 & 2249$^{**}$ & 2803$^{**}$ & - & - & 694 & 2975 & 1722 & 3828 \\
36 & 2493$^{*}$ & 2907 & - & - & 710 & 733 & 525 & 2133 \\
37 & 2494 & 2907 & - & - & 710 & 1890 & 3900 & 1691 \\
38 & 2494 & 2907 & - & - & 710 & 1891 & 831 & 1958 \\
39 & 2581$^+$ & 2584 & - & - & 710 & 3037 & 1538 & 2521 \\
40 & 2582$^+$ & 2584 & - & - & 710 & 3038 & 826 & 2225 \\
41 & 2803$^{**}$ & 3119 & - & - & 710 & 3145 & 2078 & 2734 \\
42 & 2837$^*$ & 3767 & - & - & 710 & 3146 & 2208 & 1592 \\
43 & 2838 & 3768 & - & - & 733 & 1890 & 3914 & 557 \\
44 & 2839 & 3769 & - & - & 733 & 3037 & 1838 & 3442 \\
45 & 2842$^*$ & 3772 & - & - & 733 & 3038 & 3230 & 939 \\
46 & 2875$^*$ & 2945 & - & - & 733 & 3145 & 1519 & 785 \\
47 & 2875 & 2946 & - & - & 734 & 1890 & 2927 & 2645 \\
48 & 3348$^*$ & 3377 & - & - & 734 & 3038 & 1024 & 2151 \\
49 & 3579$^{**}$ & 3850 & - & - & 847 & 3912 & 1855 & 3704 \\
50 & - & - & - & - & 874 & 876 &  375 & 3836 \\
51 & - & - & - & - & 875 & 1611 & 2626 & 54 \\
52 & - & - & - & - & 876 & 1611 & 1493 & 2839 \\ 
53 & - & - & - & - & 1032 & 1437 & 2158 & 3728 \\
54 & - & - & - & - & 1032 & 3579 & 1389 & 2943 \\
55 & - & - & - & - & 1032 & 3580 & 100 & 2045 \\
56 & - & - & - & - & 1032 & 3850 & 2130 & 2899 \\
57 & - & - & - & - & 1032 & 3851 & 1214 & 480 \\
58 & - & - & - & - & 1033 & 1438 & 1162 & 1808 \\
59 & - & - & - & - & 1033 & 3579 & 981 & 1120 \\
60 & - & - & - & - & 1033 & 3850 & 2492 & 1058 \\
61 & - & - & - & - & 1057 & 1103 & - & - \\
62 & - & - & - & - & 1057 & 3240 & - & - \\
63 & - & - & - & - & 1057 & 3331 & - & - \\
64 & - & - & - & - & 1069 & 3418 & - & - \\
65 & - & - & - & - & 1069 & 3419 & - & - \\
66 & - & - & - & - & 1162 & 1164 & - & - \\
67 & - & - & - & - & 1163 & 1165 & - & - \\
68 & - & - & - & - & 1261 & 2613 & - & - \\
69 & - & - & - & - & 1261 & 2614 & - & - \\ 
70 & - & - & - & - & 1262 & 2613 & - & - \\
71 & - & - & - & - & 1369 & 1394 & - & - \\
72 & - & - & - & - & 1369 & 2250 & - & - \\
73 & - & - & - & - & 1369 & 2804 & - & - \\
74 & - & - & - & - & 1369 & 3119 & - & - \\
75 & - & - & - & - & 1370 & 1393 & - & - \\
76 & - & - & - & - & 1370 & 1394 & - & - \\
77 & - & - & - & - & 1370 & 2249 & - & - \\
78 & - & - & - & - & 1370 & 2250 & - & - \\
79 & - & - & - & - & 1370 & 2803 & - & - \\
80 & - & - & - & - & 1370 & 3119 & - & - \\
81 & - & - & - & - & 1393 & 2249 &  - & - \\
82 & - & - & - & - & 1393 & 2803 &  - & - \\
83 & - & - & - & - & 1393 & 3119 & - & - \\
84 & - & - & - & - & 1394 & 2250 & - & - \\
85 & - & - & - & - & 1394 & 2804 & - & - \\
86 & - & - & - & - & 1394 & 3119 & - & - \\
87 & - & - & - & - & 1398 & 3455 & - & - \\
88 & - & - & - & - & 1399 & 3454 & - & - \\
89 & - & - & - & - & 1399 & 3455 & - & - \\
90 & - & - & - & - & 1421 & 1657 & - & - \\ 
91 & - & - & - & - & 1421 & 2385 & - & - \\
92 & - & - & - & - & 1437 & 3579 & - & - \\
93 & - & - & - & - & 1437 & 3580 & - & - \\ 
94 & - & - & - & - & 1437 & 3851 & - & - \\
95 & - & - & - & - & 1890 & 3038 & - & - \\
96 & - & - & - & - & 1890 & 3145 & - & - \\
97 & - & - & - & - & 1891 & 2177 & - & - \\
98 & - & - & - & - & 1891 & 2178 & - & - \\
99 & - & - & - & - & 1891 & 3958 & - & - \\
100 & - & - & - & - & 2041 & 2046 & - & - \\
101 & - & - & - & - & 2059 & 2061 & - & - \\
102 & - & - & - & - & 2141 & 2143 & - & - \\
103 & - & - & - & - & 2249 & 2803 & - & - \\
104 & - & - & - & - & 2250 & 2803 & - & - \\
105 & - & - & - & - & 2250 & 2804 & - & - \\
106 & - & - & - & - & 2250 & 3119 & - & - \\
107 & - & - & - & - & 2303 & 3240 & - & - \\
108 & - & - & - & - & 2303 & 3958 & - & - \\
109 & - & - & - & - & 2385 & 3958 & - & - \\ 
110 & - & - & - & - & 2493 & 2907 & - & - \\
111 & - & - & - & - & 2493 & 2958 & - & - \\
112 & - & - & - & - & 2494 & 2906 & - & - \\
113 & - & - & - & - & 2494 & 2907 & - & - \\
114 & - & - & - & - & 2494 & 2957 & - & - \\
115 & - & - & - & - & 2495 & 2906 & - & - \\
116 & - & - & - & - & 2495 & 2957 & - & - \\
117 & - & - & - & - & 2581 & 2583 & - & - \\
118 & - & - & - & - & 2581 & 2584 & - & - \\
119 & - & - & - & - & 2582 & 2584 & - & - \\
120 & - & - & - & - & 2655 & 3912 & - & - \\
121 & - & - & - & - & 2803 & 3118 & - & - \\
122 & - & - & - & - & 2803 & 3119 & - & - \\
123 & - & - & - & - & 2804 & 3119 & - & - \\
124 & - & - & - & - & 2837 & 3767 & - & - \\
125 & - & - & - & - & 2837 & 3768 & - & - \\
126 & - & - & - & - & 2838 & 3767 & - & - \\
127 & - & - & - & - & 2838 & 3768 & - & - \\ 
128 & - & - & - & - & 2838 & 3769 & - & - \\
129 & - & - & - & - & 2839 & 3769 & - & - \\
130 & - & - & - & - & 2839 & 3770 & - & - \\
131 & - & - & - & - & 2840 & 3770 & - & - \\
132 & - & - & - & - & 2841 & 3770 & - & - \\
133 & - & - & - & - & 2842 & 3773 & - & - \\
134 & - & - & - & - & 2842 & 3772 & - & - \\
135 & - & - & - & - & 2842 & 3773 & - & - \\
136 & - & - & - & - & 2874 & 2946 & - & - \\
137 & - & - & - & - & 2875 & 2945 & - & - \\
138 & - & - & - & - & 2875 & 2946 & - & - \\
139 & - & - & - & - & 2876 & 2945 & - & - \\
140 & - & - & - & - & 2906 & 2957 & - & - \\
141 & - & - & - & - & 2907 & 2958 & - & - \\
142 & - & - & - & - & 3008 & 3011 & - & - \\
143 & - & - & - & - & 3037 & 3145 & - & - \\
144 & - & - & - & - & 3038 & 3145 & - & - \\
145 & - & - & - & - & 3241 & 3591 & - & - \\
146 & - & - & - & - & 3241 & 3957 & - & - \\ 
147 & - & - & - & - & 3348 & 3377 & - & - \\
148 & - & - & - & - & 3579 & 3850 & - & - \\
149 & - & - & - & - & 3579 & 3851 &  - & - \\
150 & - & - & - & - & 3580 & 3851 & - & - \\
151 & - & - & - & - & 3591 & 3957 & - & - \\
152 & - & - & - & - & 3764 & 3958 & - & - \\
153 & - & - & - & - & 4007 & 4011 & - & - \\

\hline
\bottomrule
\caption{\label{tab:table1}
Table showing the list of pair of monomers which constitute the  CLs for {\em C. Crecentus}, these CLs are 
used as an input to the  simulation by constraining these monomers to be at a distance $a$ from each other.
The first monomer with label $1$ and the last monomer labelled $4017$ are linked together because the DNA is a ring polymer. 
}
\end{longtable*}

\null\pagebreak \null\pagebreak

\clearpage

\section{ Positional Correlations}
\begin{figure}[!hbt]
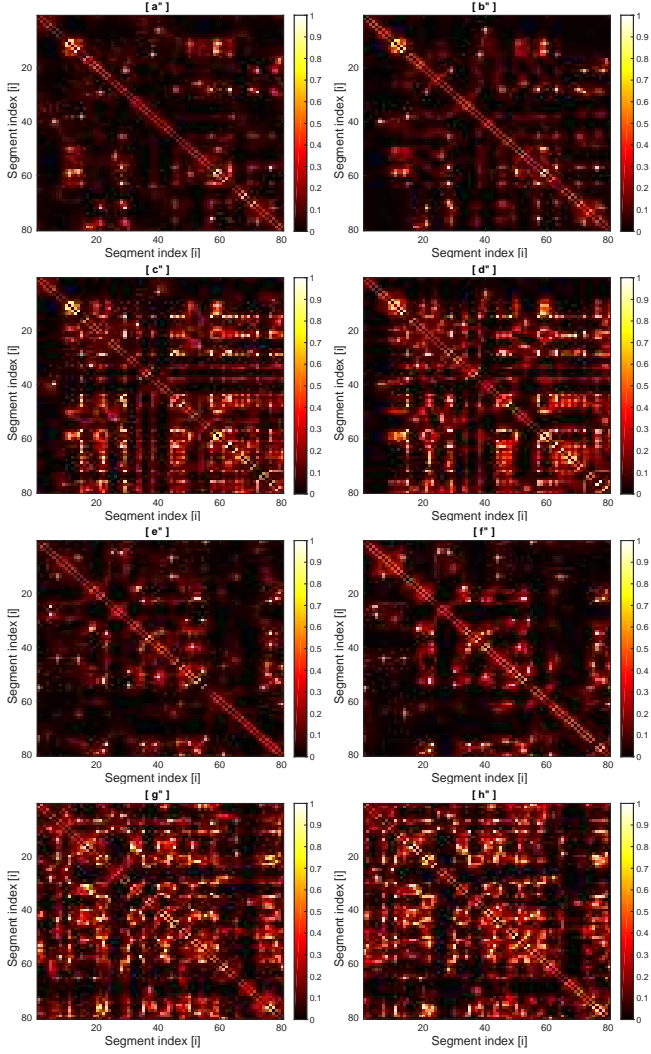

\includegraphics[width=0.49\columnwidth]{bio_cc_i5_cp1_cm_dom_corr1_l1.eps}
\includegraphics[width=0.49\columnwidth]{bio_cc_i7_cp1_cm_dom_corr1_l1.eps} \\
\includegraphics[width=0.49\columnwidth]{bio_cc_i5_cp4_cm_dom_corr1_l1.eps}
\includegraphics[width=0.49\columnwidth]{bio_cc_i7_cp4_cm_dom_corr1_l1.eps} \\
\includegraphics[width=0.49\columnwidth]{r_bio_cc_i5_cp1_cm_dom_corr1_l1.eps}
\includegraphics[width=0.49\columnwidth]{r_bio_cc_i7_cp1_cm_dom_corr1_l1.eps} \\
\includegraphics[width=0.49\columnwidth]{r_bio_cc_i5_cp4_cm_dom_corr1_l1.eps}
\includegraphics[width=0.49\columnwidth]{r_bio_cc_i7_cp4_cm_dom_corr1_l1.eps} \\
\hfill
\caption{\label{suppfig2}
The set of 8 colormaps give the probability of finding the center of mass of segment $i$ and
segment $j$ of rings polymers within a distance cutoff of $R^C = 5a$. The {\em C. Crescentus} bacteria ring polymer with $4017$ monomers
is considered to be a set of 80 segments with $50$ monomers in each segment. These data provide more structural information
of the local arrangements of polymer segments than a stardard pair correlation function $g(r)$ data. Figure (a"),(b") correspond
different runs of a  polymer chain with cross links (CL), the locations of CLs  correspond to BC-1.
Each run stars with different initial configuration. Subplots (c"),(d") similarly are for different runs with
BC-2. The data shown in (e"),(f") are for different runs with  RC-1, and (g"),(h") for runs with
RC-2. This figure gives data from extra runs, corresponding to that of Fig. 2 of main manuscript.}
\end{figure}

\null\pagebreak

\section{ Angular Correlations}
\vskip0.8cm
\begin{figure}[!hbt]
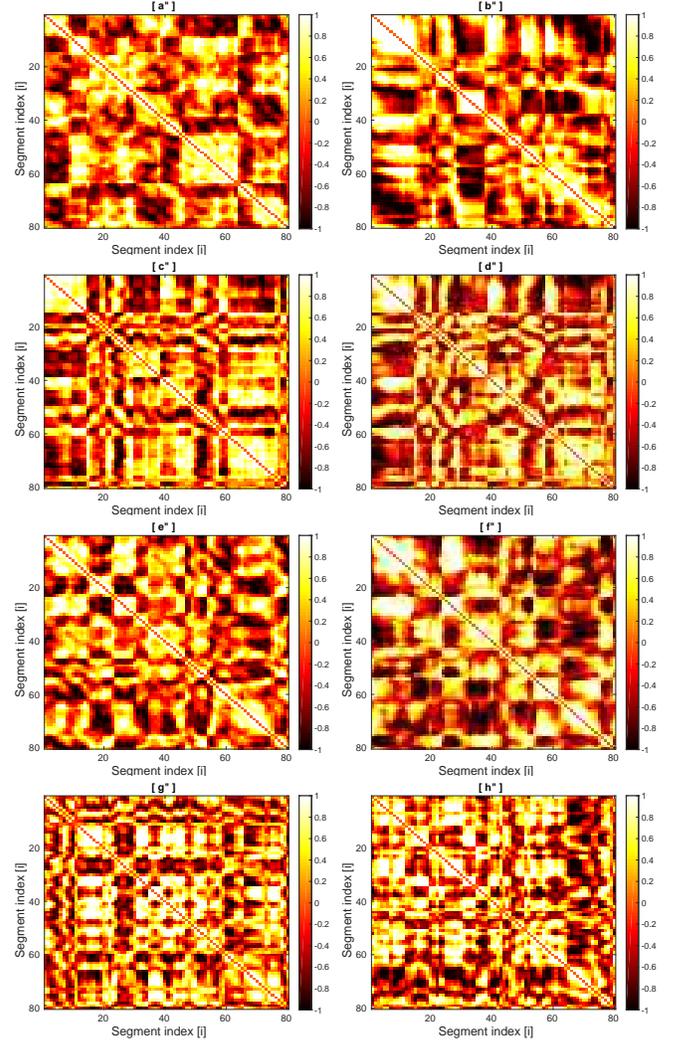

\includegraphics[width=0.49\columnwidth]{bio_cc_i5_cp1_near_contacts1_l1.eps}
\includegraphics[width=0.49\columnwidth]{bio_cc_i7_cp1_near_contacts1_l1.eps}
\hfill
\includegraphics[width=0.49\columnwidth]{bio_cc_i5_cp4_near_contacts1_l1.eps}
\includegraphics[width=0.49\columnwidth]{bio_cc_i7_cp4_near_contacts1_l1.eps}
\hfill
\includegraphics[width=0.49\columnwidth]{r_bio_cc_i5_cp1_near_contacts1_l1.eps}
\includegraphics[width=0.49\columnwidth]{r_bio_cc_i7_cp1_near_contacts1_l1.eps}
\hfill
\includegraphics[width=0.49\columnwidth]{r_bio_cc_i5_cp4_near_contacts1_l1.eps}
\includegraphics[width=0.49\columnwidth]{r_bio_cc_i7_cp4_near_contacts1_l1.eps}
\caption{\label{suppfig3} 
The set of 8 colormaps give the probability of finding the angular correlations between segment $i$ and
segment $j$ of a ring polymer. These data provide more structural information
of the local arrangements of polymer segments than a standard pair correlation function $g(r)$ data.
Figure (a"),(b") correspond to
different runs of a  polymer chain with cross links at positions corresponding to BC-1.
Each run starts with a different initial configuration. Subplots (c"),(d") similarly are for different runs with
BC-2. The data shown in (e"),(f") are for different runs with  RC-1, and (g"),(h") for runs with
RC-2. This figure gives data from extra runs and correspond to Fig.3 of main manuscript.}
\end{figure}

\clearpage
\section{ Energy}
\begin{figure}[!hbt]
\includegraphics[width=0.70\columnwidth]{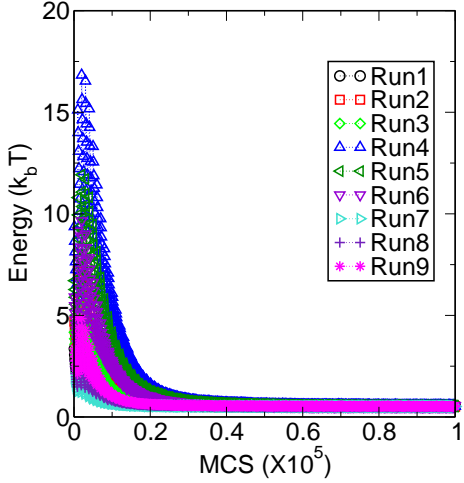}
\caption{\label{suppfig1}
Plot shows the potential energy of the polymer chain with CLs as it relaxes to the same value of the potential energy.Each of the 
$9$ runs start with independent initial configuration of the polymer.}

\end{figure}

\section{Radial distribution of Segments}
\begin{figure}[!hbt]
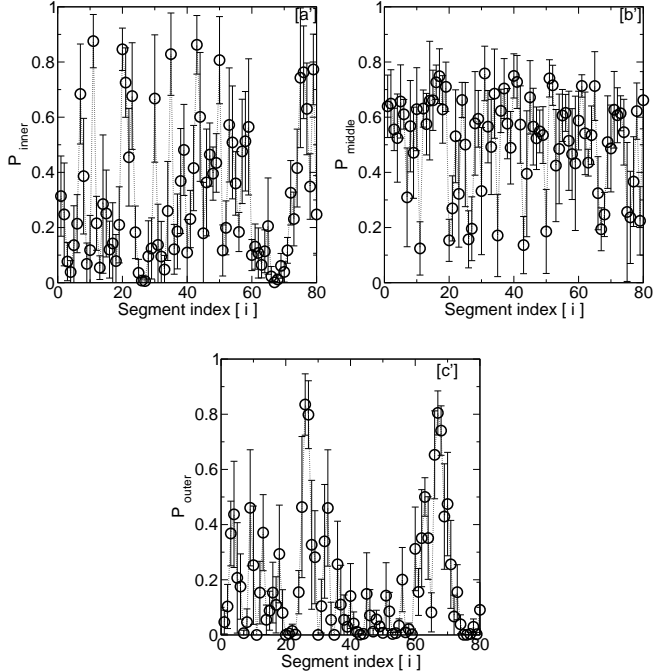

\includegraphics[width=0.49\columnwidth]{ran_cm_contact4_inner.eps}
\includegraphics[width=0.49\columnwidth]{ran_cm_contact4_middle.eps} 
\vskip0.4cm
\includegraphics[width=0.49\columnwidth]{ran_cm_contact4_outer.eps}
\caption{\label{suppfig5}
Plot shows the probability $P_{inner},\,P_{middle}, \,P_{outer}$ of finding a segment $i$ in the inner, middle
and outer regions, respectively, of the polymer for RC-2.}

\end{figure}

\section{Experimental contact-maps}
\begin{figure}[!hbt]
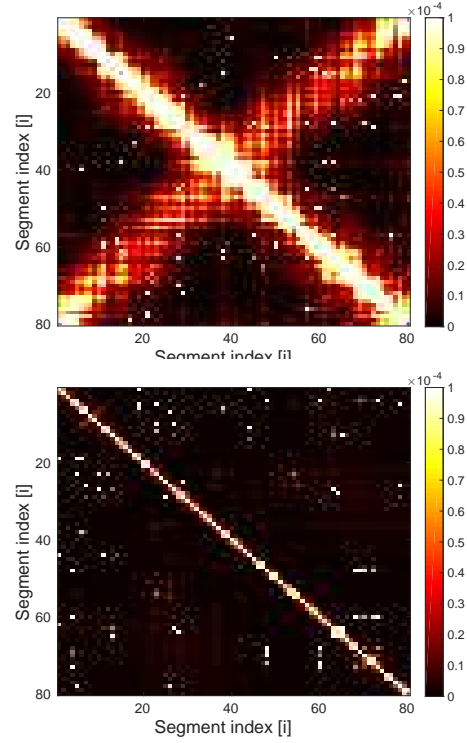

\includegraphics[width=0.70\columnwidth]{matrix_caul_manip.eps} \\
\includegraphics[width=0.70\columnwidth]{matrix_ecoli_manip.eps}
\caption{\label{suppfig4}
Figure shows colormap obtained from the analysis of Hi-C data and the color represent the 
interaction frequencies in the experiment. Top colormap is for bacteria {\em C.Crescentus}
and the bottom colormap is for the {\em E.Coli}.}
\end{figure}

\section{Reasons for differences between colormaps of Fig.2 and Fig.3}
Comparing the colormaps of positional and angular correlations in Fig.S1 c",d" with Fig.S2 c",d"
(or equivalently Fig.\ref{fig2} c,d with Fig.\ref{fig3} c,d) for BC-2, 
we observe relatively large patches
of bright pixels along the diagonal in Fig.\ref{suppfig3} c",d".  The reason for this difference is as follows.
the cutoff distance chosen for the calculations in Fig.\ref{suppfig2} c",d" is $R^C =5a$; this is nearly equal
to the value of $R_g$  of the $50$-monomer segments,
viz.,$R_g \sim {50}^{0.6}a/\sqrt{6} \approx 4.2a$.
Thus fewer  segments can be accommodated within distance $R^C$ from the CM of a particular segment,
resulting in small bright patches. In contrast, for angular correlations, we just calculate whether two
particular segments lie on the same or opposite side with respect to CM of coil.
The relatively large patches of bright pixels in Fig.S2 c",d" is a consequence of choice of binary values
of $\delta^{ij}_\theta$ for a particlar microstate.
The patterns would be more fuzzy if chose to plot $\langle \cos(\theta^{ij}) \rangle$ instead
of $\langle \delta^{ij}_\theta \rangle$.
\clearpage

\section{Construction of 2-D map}
To construct the 2-D map, we assumed a spherical globular structure of the DNA-coils
and then we use a 3-step procedure in conjunction. Firstly, we know from 
Fig.1 which segments are to be found primarily in the inner, middle and peripheral region of a globule.
Secondly and thirdly, we use the information about which segments are to be found on the radially opposite hemispheres 
(Fig.3), in tandem with the data about positional correlation given in Fig.2. We use 
this collective quantitative information to distribute  the segments within the schematic diagram of the globule.
\section{Snapshot from simulation}
\begin{figure}[!hbt]
\includegraphics[width=0.6\columnwidth]{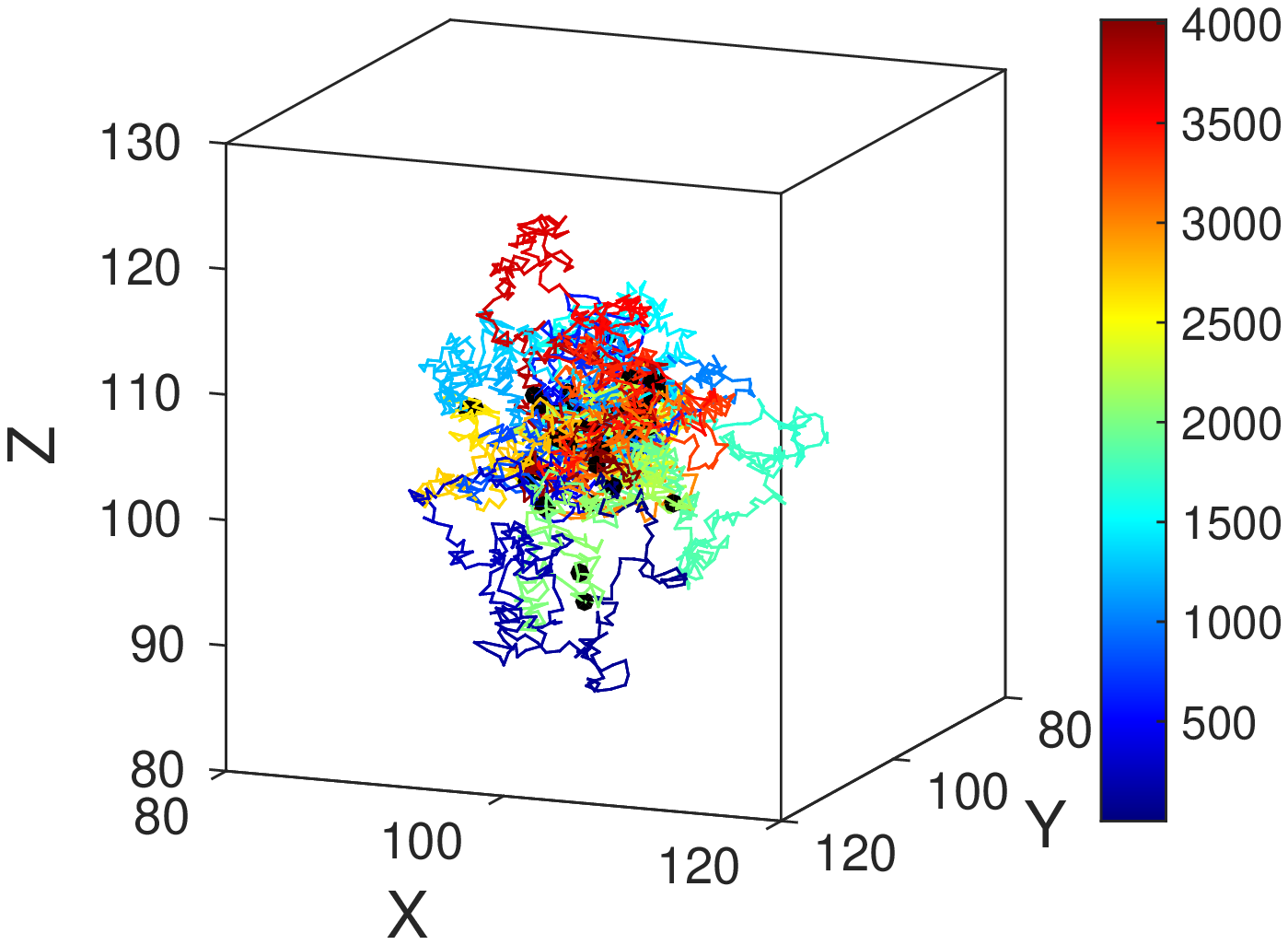} \\
\includegraphics[width=0.6\columnwidth]{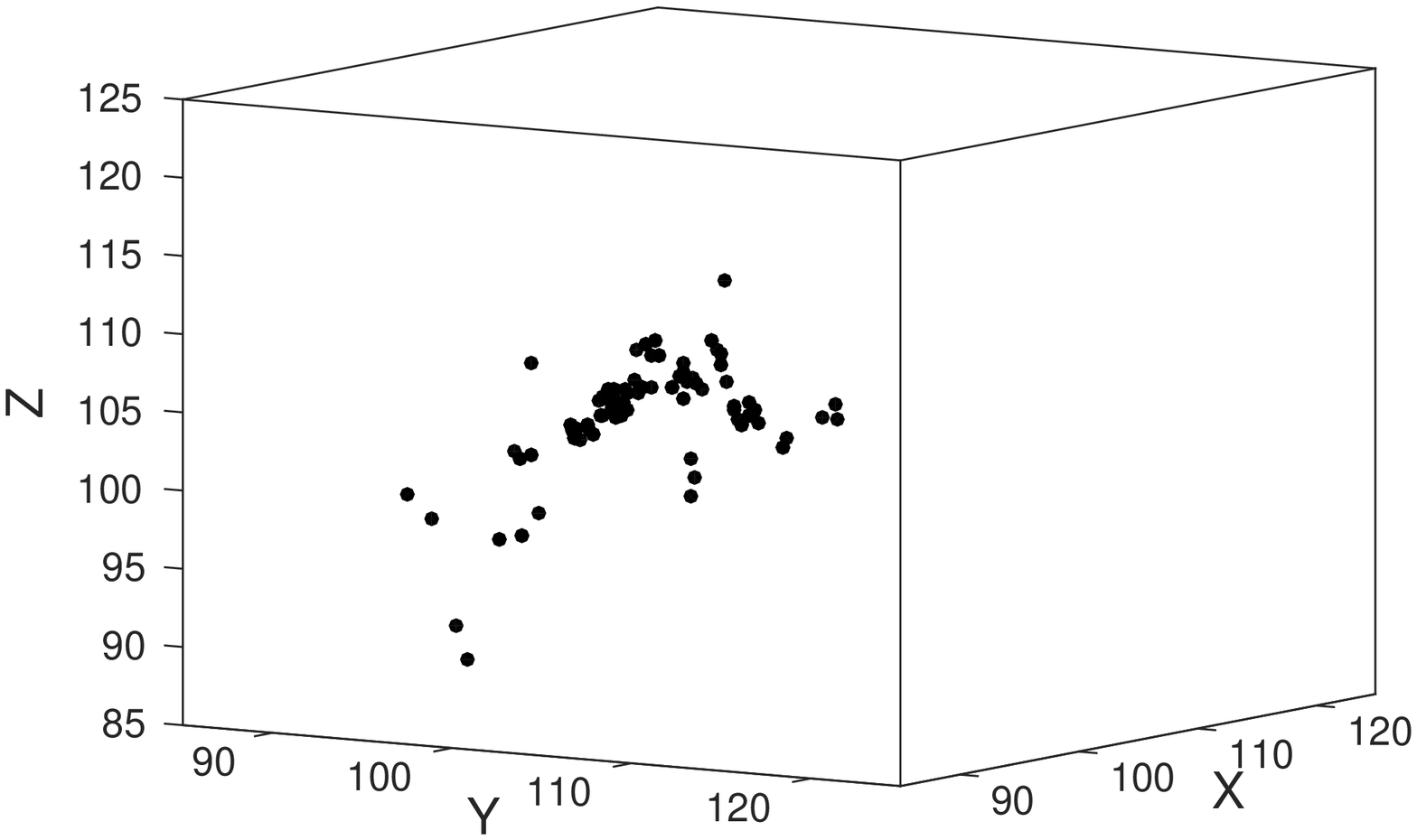} \\
\includegraphics[width=0.6\columnwidth]{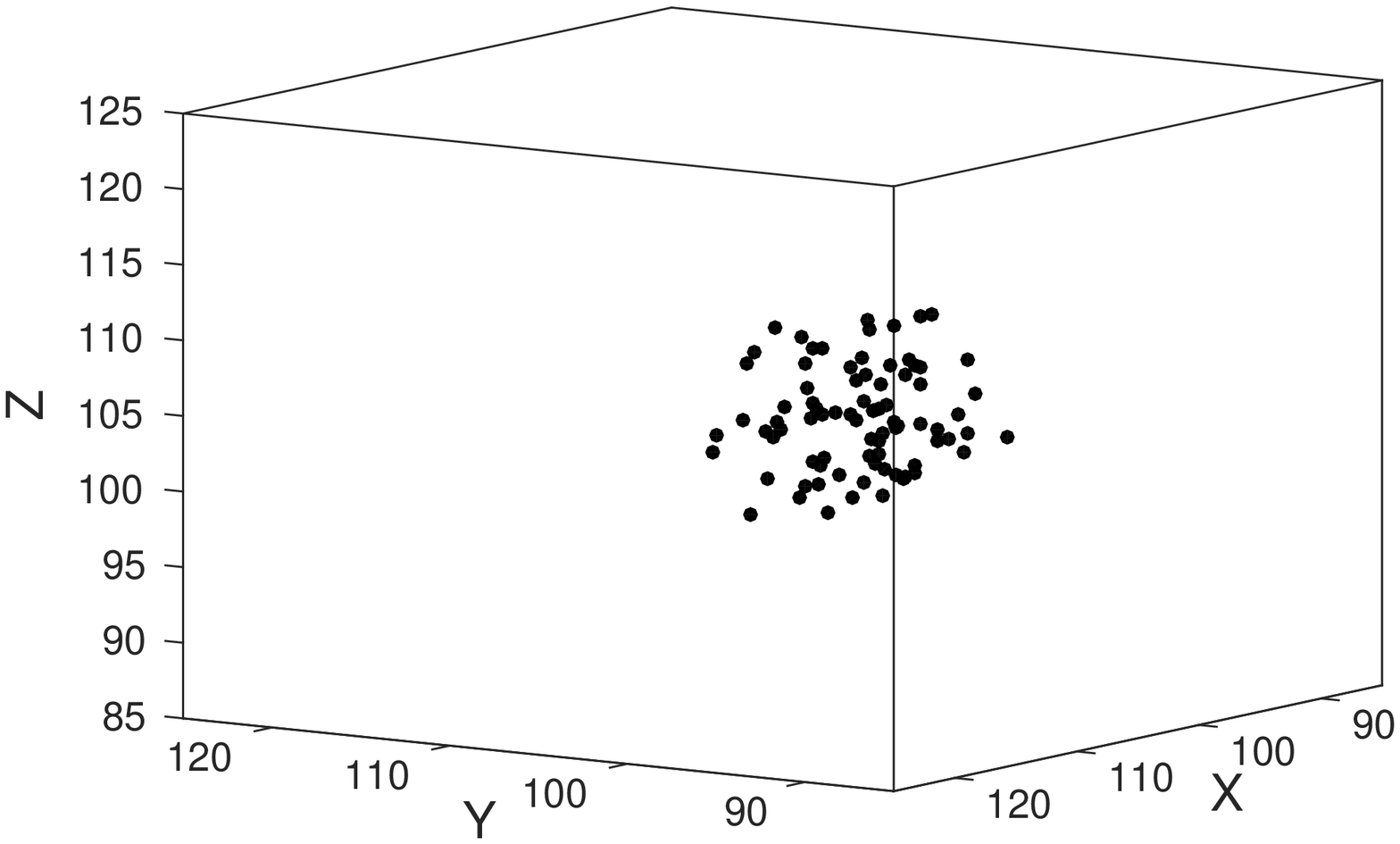} 
\caption{\label{Suppsnapshot}
Representative snaphots of the polymer with BC-2(top) CLs form our simulation. Black circles show the
positions of the CLs. The monomers are colored from blue to red along the chain contour. The positions of CLs for BC-2, RC-2 is also shown in middle, bottom figures
where the polymer is not shown for better visualization.
From the plots we see that CLs are clustered in space for BC-2 and are scattered for RC-2.
}
\end{figure}
\clearpage

\end{document}